\documentclass[letterpaper]{article} 
\usepackage{aaai25}  
\usepackage{times}  
\usepackage{helvet}  
\usepackage{courier}  
\usepackage[hyphens]{url}  
\usepackage{graphicx} 
\usepackage{natbib}  
\usepackage{caption} 
\usepackage{algorithm}
\usepackage{algorithmic}

\usepackage{newfloat}
\usepackage{listings}
\DeclareCaptionStyle{ruled}{labelfont=normalfont,labelsep=colon,strut=off} 
\floatstyle{ruled}
\newfloat{listing}{tb}{lst}{}
\floatname{listing}{Listing}
\usepackage{acronym}
\acrodef{ANN}[ANN]{Artificial Neural Network}
\acrodef{PS}[program synthesis]{program synthesis}
\acrodef{ARC}[ARC]{Abstraction and Reasoning Challenge}
\acrodef{LLM}[LLM]{Large Language Model}
\acrodef{PBE}[PBE]{Programming-By-Example}
\acrodef{DSL}[DSL]{Domain-Specific Language}
\acrodef{CB}[CrossBeam]{CrossBeam}
\acrodef{LB}[LambdaBeam]{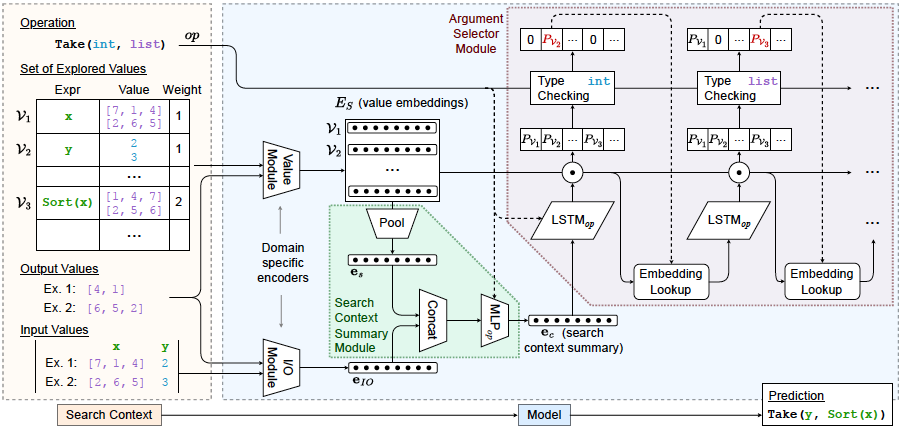}
\acrodef{AB}[AbstractBeam]{AbstractBeam}
\acrodef{io}[I/O]{Input-Output}
\acrodef{LLM}[LLM]{Large Language Model}
\usepackage{amsmath} 
\usepackage{xcolor}
\usepackage{listings}
\usepackage{pgfplots}
\pgfplotsset{compat=newest}
\pgfplotsset{scaled y ticks=false}
\usepgfplotslibrary{groupplots}
\usepgfplotslibrary{dateplot}
\usepackage{subcaption}
\usepackage{multirow}
\usepackage{cleveref} 
\usepackage{pdfpages}

\nocopyright
\title{AbstractBeam: Enhancing Bottom-Up Program Synthesis Using Library Learning}
\author{
    Janis Zenkner\equalcontrib\textsuperscript{\rm 1},
    Lukas Dierkes\equalcontrib\textsuperscript{\rm 1},
    Tobias Sesterhenn\textsuperscript{\rm 1},
    Christian Bartelt\textsuperscript{\rm 1}
}
\affiliations{
    \textsuperscript{\rm 1} Institute for Enterprise Systems

    University of Mannheim\\
    Mannheim, 68161\\
    janis.zenkner@uni-mannheim.de
}

\begin{document}

\maketitle

%

\begin{abstract}
LambdaBeam is a state-of-the-art, execution-guided algorithm for program synthesis that utilizes higher-order functions, lambda functions, and iterative loops within a Domain-Specific Language (DSL). 
LambdaBeam generates each program from scratch but does not take advantage of the frequent recurrence of program blocks or subprograms commonly found in specific domains, such as loops for list traversal.
To address this limitation, we introduce AbstractBeam: a novel program synthesis framework designed to enhance LambdaBeam by leveraging Library Learning.
AbstractBeam identifies and integrates recurring program structures into the DSL, optimizing the synthesis process.
Our experimental evaluations demonstrate that AbstractBeam statistically significantly ($p < 0.05$) outperforms LambdaBeam in the integer list manipulation domain.
Beyond solving more tasks, AbstractBeam's program synthesis is also more efficient, requiring less time and fewer candidate programs to generate a solution.
Furthermore, our findings indicate that Library Learning effectively enhances program synthesis in domains that are not explicitly designed to showcase its advantages, thereby highlighting the broader applicability of Library Learning.
\end{abstract}
\section{Introduction}
The field of \acs{PS} aims to enable machines to automatically generate programs from a set of user-provided specifications, presenting a challenge in artificial intelligence and computer science~\cite{gulwani2017program}.
Various techniques have been developed to tackle this challenge, including Graph Neural Networks~\cite{allamanis2017learning}, Sequence-to-Sequence models~\cite{alon2020structural}, and Large Language Models~\cite{chen2021evaluating}. 
In addition to these foundational end-to-end approaches, the use of \acp{DSL} and \ac{PBE} can refine and target the automated generation of programs.

\ac{PBE} uniquely allows program specifications through \ac{io} examples, from which a synthesis tool generates the corresponding code~\cite{polozov2015flashmeta}.
Meanwhile, \acp{DSL} are specialized languages designed to address specific problems within particular domains, offering concise syntax and operations optimized for targeted tasks~\cite{van2000domain}.

A straightforward approach to solve a \ac{PBE} task is to generate and execute all possible combinations of operations within a DSL. 
However, the number of combinations grows exponentially with program length, leading to an infeasible combinatorial explosion as the depth and breadth of the search space expand. 
This search space can be visualized as a decision tree, where longer programs increase depth and more complex \acp{DSL} increase breadth.
Consequently, the trial-and-error approach becomes impractical for complex tasks.

To address this, many \ac{DSL}-based \acs{PS} approaches apply combinatorial searches guided by \acp{ANN} or \acp{LLM} to navigate the search space more efficiently~\cite{balog2016deepcoder, shi2019frangel, barke2020just, shi2022tf, grand2023lilo}.
However, the exponential growth in search space remains a critical challenge, often limiting the effectiveness of \ac{DSL}-based \acs{PS}.

Additionally, programming is an inherently complex cognitive task, combining logic, creativity, and systematic problem-solving~\cite{pea1984cognitive}.
Human programmers often employ a \textit{bottom-up approach}, crafting small, self-contained functions to address specific aspects of a larger problem and then combining these functions to construct a complete software solution~\cite{cormen2001advanced}.
Another strategy to reduce complexity and development time is the use of pre-built software libraries. 
For example, in the domain of Machine Learning, practitioners use model architectures from PyTorch to streamline their development process~\cite{stevens2020deep}.

Inspired by human software development, approaches like DreamCoder~\cite{ellis2021dreamcoder} incorporate Library Learning into a top-down synthesis framework. 
Library Learning identifies and extracts frequently occurring program patterns or subprograms, creating reusable components for more efficient code generation which are added to the initial \ac{DSL}. 
DreamCoder integrates this procedure into a wake-sleep cycle: during the wake phase, a simple Bayesian top-down algorithm generates programs; during the sleep phase, repeating subprograms are identified and added to the \ac{DSL}. 
By abstracting subprograms containing multiple operations and constants, the length of synthesized programs is reduced, decreasing the search depth and improving efficiency.

\acs{LB}~\cite{shi2024lambdabeam} focuses on narrowing the search breadth using an execution-guided bottom-up search. During bottom-up synthesis, subprograms are executed, and subsequent searches are informed by these intermediate program states, providing more fine-grained guidance.
Additionally, \acs{LB} incorporates higher-order functions, enhancing its capability to handle complex synthesis tasks.

While \acs{LB} is efficient in terms of the number of programs evaluated, it does not utilize Library Learning, leading to a time-consuming generation of these programs.  
Conversely, DreamCoder excels with Library Learning and abstractions but lacks the fine-grained contextual information available during execution-guided synthesis.
Furthermore, the domains DreamCoder was evaluated on contain multiple task clusters that differ only by a single parameter (e.g., \textit{add-k with k=1}, \textit{add-k with k=2}, etc.). 
This makes it easy to find and apply abstractions, creating an environment that favors Library Learning approaches but overlooks scalability and generalization challenges.

This situation presents an opportunity for a unified approach that combines the strengths of execution-guided bottom-up synthesis and Library Learning.
We hypothesize that enhancing \acs{LB} with Library Learning will improve performance by reducing program length and search depth. 
Additionally, we believe that abstractions can be found in domains not specifically designed to contain program repetitions and that these abstractions do not generalize beyond the domain unless they accurately reflect it.

Driven by these hypotheses, we propose a new framework, \acs{AB}, which leverages \acs{LB}'s search policy and a Library Learning framework inspired by DreamCoder. 
\acs{AB} can define, use, and reason about operations and abstractions that may include lambda functions, iterative loops, and higher-order functions.
We demonstrate the effectiveness of \acs{AB} in the \acs{LB} list manipulation domain exploring its applicability in a domain not specifically designed to showcase program repetitions, providing insights into the scalability and generalization capabilities of \acs{AB}.
Our \acs{AB} code is available in the supplementary materials\footnote{The code will be made publicly available after publication.}.

\section{Background}
\label{sec:Background}
\subsection{Programming By Example}
\ac{PBE} is a technique that allows users to describe programs through \ac{io} examples rather than natural language specifications. 
The objective in \ac{PBE} tasks is to find a program $\mathcal{P} \: \epsilon \: \mathcal{L}$ that correctly maps all inputs ${\mathcal{I} = \{I_{1}, ..., I_{N}\}}$ to their respective outputs $\mathcal{O} = \{O_{1}, ..., O_{N}\}$. 
Here, $\mathcal{L}$ represents the space of all possible programs, defined by a \ac{DSL} that includes constants, input variables, and operations. 
Operations can be arbitrarily nested and be applied to constants and input variables. 
Exemplary tasks can be found in the supplementary materials.

\subsection{\texorpdfstring{$\mathbf{\lambda}$}{lambda}-Calculus}
The lambda calculus~\cite{church1985calculi} is a formal system in mathematical logic that can simulate any Turing Machine. 
In lambda calculus, terms are defined as variables, function applications, and lambda abstractions. 
Lambda abstractions create new functions with a specific set of variables that define their scope and the function's body.
Unlike traditional lambda functions, which introduce variables one at a time through a process known as "currying," we allow lambda abstractions to introduce multiple variables simultaneously.
Moreover, we incorporate manually designed, domain-specific primitives, such as \textit{Add}, \textit{Sort}, or \textit{Map}, to extend the lambda calculus.
Additionally, we employ simply-typed lambda calculus~\cite{pierce2002types} aiding in data type management and function validation.

\subsection{\acs{LB}}
\acs{LB} is a program synthesis framework that utilizes a bottom-up approach, efficiently generating programs by recombining previously explored programs termed "values".
These values are stored in a set that initially only contains input variables and constants.
For each step, \acs{AB} iterates over the available operations in the \ac{DSL} and selects the arguments for the next operation from this set, based on the operation's arity, i.e., the number of arguments it requires.
A neural module assesses the search context, which includes the \ac{io} examples, the current operation $f$, and the previously explored values. 
This module outputs a probability distribution over the set of explored values, guiding a beam search algorithm in selecting the most appropriate arguments $a_i$ for the operation $f$.
Once all necessary arguments are selected, the function $f(a_0, \ldots, a_{\text{arity}})$ is executed, and its output is added to the set of explored values.
This process continues until the algorithm successfully synthesizes a program that meets the specified requirements or until the search times out.

The bottom-up generation allows \acs{LB} to construct programs incrementally, refining its search based on intermediate execution results.
Deterministic beam search can stall if newly generated values duplicate previously encountered ones, leaving the set of explored values unchanged.
To avoid this, the UniqueRandomizer~\cite{shi2020incremental} methodically selects unique samples and records them in a specialized data structure, preventing the reselection of previously chosen values. 
Moreover, the search process includes periodic restarts to counteract difficulties in recovering from ineffective search paths.
These restarts help refresh the search space, though at the cost of discarding previously explored values.
Further details on \acs{LB}'s learning algorithm are available in the supplementary materials.

\subsection{Stitch}
Stitch~\cite{bowers2023top} implements an automated synthesis method that extracts abstractions from a corpus of programs. It begins with an initial \ac{DSL} and employs a top-down search method to identify abstractions that compress the program corpus.

One challenge in this approach is that programs in the corpus may not be in normal form, meaning semantically equivalent subprograms may appear in different syntactical forms, e.g., $t_1(x) = x + 1$ and $t_2(x) = 1 + x$, complicating their identification. 
Transforming programs into normal form using defined rules can be impractical due to the increasing complexity and frequency of rule applications, akin to the search space blow-up in program generation.

To address this, Stitch forgoes predefined rewrite rules and instead employs a synthesis algorithm that generates programs not to solve tasks but to compress the program corpus by identifying common patterns.
This approach sacrifices some completeness in compression for a more efficient identification process.

Stitch begins with an empty abstraction and sequentially adds operations from the \acs{DSL}. 
Initially, arguments in abstractions are placeholders, allowing them to be filled with new operations, variables, or constants in subsequent iterations. 
To enhance generality, Stitch supports partial abstractions, which introduce new variables called "holes" (\texttt{??}) to fill placeholders.
\begin{equation*}
    \begin{alignedat}[b]{1}
    f_1(x) &= \text{Add}(3, \text{Subtract}(\text{Multiply}(2, 3), 1))\\
    f_2(x) &= \text{Add}(3, \text{Subtract}(\text{Add}(1, 2), \text{Square}(4)))\\
    a_{\text{par}}(x) &= \text{Add}(3\:,\:??)
\end{alignedat}
\end{equation*}
For example, given a program corpus with $f_1(x)$ and $f_2(x)$ and a partial abstraction $a_{\text{par}}(x)$, Stitch iterates over \ac{DSL} operations to compare and extend this abstraction (e.g., $a_{\text{par}}(x) = \text{Add}(3\:, \text{Subtract}(\cdot, \cdot))$.
An exhaustive top-down search over all operations is infeasible due to the breadth and depth of the search tree.
Instead, Stitch selects actions that optimize the compression of the program corpus.
It prioritizes operations that maximize the product of an abstraction's length and frequency within the corpus: If a newly added operation's contribution is lower than the current abstraction's, the entire branch can be pruned, leading to significant gains in processing speed and memory efficiency.

\section{AbstractBeam}
\label{sec:Methods}
To demonstrate the potential that Library Learning brings to \acs{PS} approaches, consider a \ac{DSL} and a target program $p(l)$ that doubles all even elements in a list $l$ and drops all odd ones.\footnote{This illustrative syntax differs from the programs generated by \acs{LB} and \acs{AB}.}
\begin{equation*}
    \begin{alignedat}[b]{1}
    \text{DSL} &= [\text{IsEven}(x),\:\text{Double}(x),\:\text{If}(x, \text{then}, \text{else}),\\ 
    &\quad\: \text{Loop}(l, \text{start}, \text{stop}),\:\text{Len}(l)]\\
    \text{constans} &= [0,\:1,\:2,\:3]\\
    p(l) &= \text{Loop}(l, 0, \text{Len}(l))\\
    &\quad (\text{If}(\text{IsEven}(x), \text{Double}(x), \text{Drop}(x)))
    \end{alignedat}
\end{equation*}

Using an exhaustive search up to a maximum length of 8, more than $|DSL|^{\text{program length}}=9^8 > 43$ million candidate programs have to be tested to find the solution. 
However, if \acs{LB} consistently suggests the correct operation, only a single candidate program of weight 8 would be generated.
Realistically, \acs{LB} may struggle to determine the start and end parameters of the loop operation.
Assuming five options are available for both parameters (the constants and $\text{Len}(l)$),  there are $5^2 = 25$ candidate programs to execute to find the solution, which is over six orders of magnitude fewer than with the exhaustive approach.
Library Learning further optimizes this process by identifying and reusing common abstractions. 
Suppose that \acs{AB} has already found an abstraction $a(l)$ that performs a loop over the entire list.
With the same assumption as for \acs{LB}, \acs{AB} suggests a single candidate program $p_{\text{AB}}(l)$ of length 5 that solves the task.
\begin{equation*}
    \begin{alignedat}[b]{1}
    a(l) & = \text{Loop}(l, 0, \text{Len}(l))\\
    p_{\text{AB}}(l) & = a(l)(\text{If}(\text{IsEven}(x), \text{Double}(x), \text{Drop}(x)))
    \end{alignedat}
\end{equation*}
This example illustrates the efficiency of \acs{LB} in finding a solution with minimal candidate programs.
\acs{AB} enhances this capability and accelerates solution discovery by shortening program length through the use of abstractions.
Consequently, \acs{AB} improves the runtime efficiency of \acs{LB}, offering a more streamlined approach to program synthesis.

\subsection{\acs{AB} Framework}
\begin{figure}[t]
    \centering
    \includegraphics[width=\linewidth]{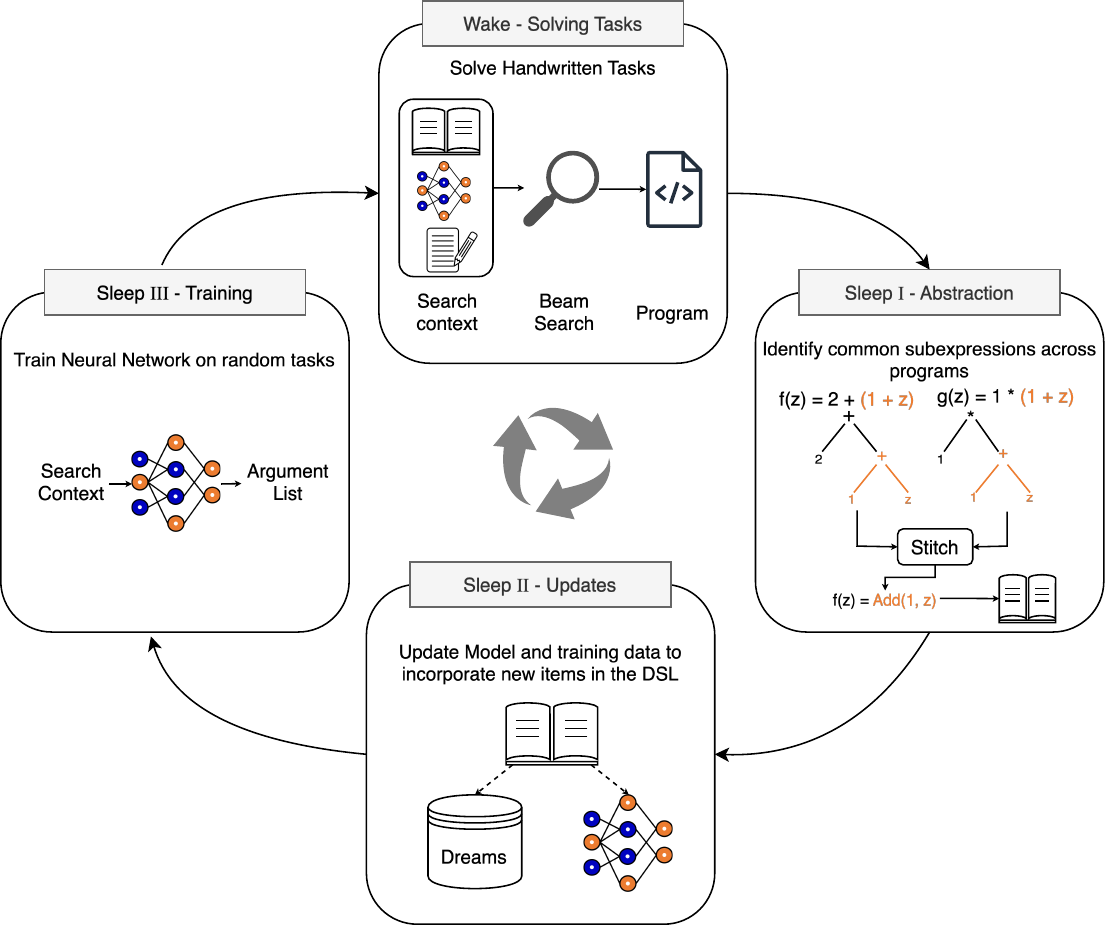}
    \caption[Overview AbstractBeam]{Systematic overview of wake-sleep cylce of the \acs{AB} framework. During the wake phase, programs are generated to solve tasks. Solutions are then used in the sleep phase to find abstractions. Finally, the model architecture and the training data are adapted to newfound abstractions.}
    \label{figure:abstractbeam}
\end{figure}

\acs{AB} enhances program synthesis by incorporating a learning cycle, similar to DreamCoder~\cite{ellis2021dreamcoder}. 
As illustrated in Figure~\ref{figure:abstractbeam}, this cycle consists of two main phases.
During the wake phase, the search policy attempts to solve as many tasks as possible using the current \ac{DSL}.
In the sleep phase, the compression algorithm leverages the task solutions from the wake phase to add abstractions to the \ac{DSL}, which aids in solving more tasks in the next iteration and consequently in discovering more abstractions.

\paragraph{Wake Phase}
In the wake phase, a bottom-up search policy inspired by \acs{LB} is employed to generate programs that solve the training tasks based on given \ac{io} pairs.
The architecture of the neural model comprises four key components: the \acs{io} Module, the Value Model, the Search-Context Summary Module, and the Argument Selector Module.
The \acs{io} Module, based on a MLP, processes task examples into a vector.
The Value Model embeds explored values and their weights with another MLP.
These embeddings are combined into a vector representation by the Search-Context Summary Module. 
Finally, the Argument Selector Module, based on an LSTM architecture, predicts a probability distribution over the set of possible next steps, i.e., which arguments to use for the current operation.
This probability distribution is used as a scoring function for a beam search algorithm to select the most promising arguments.
Details on the model architecture can be found in the supplementary materials.

Unlike \acs{LB}, where techniques such as Unique Randomizer and random restarts are typically used only during testing, \acs{AB} applies these methods during the wake phase as well. 
This ensures a sufficient diversity of programs for the Library Learning algorithm to work with, increasing the likelihood of discovering useful abstractions.

\paragraph{Sleep Phase}
In contrast to \acs{LB}, where the focus is purely on program synthesis, \acs{AB} leverages Stitch~\cite{bowers2023top} to compress and optimize the \ac{DSL}. 
This compression is done during the sleep phase which is divided into two parts: abstraction identification and model adaption and training.

The first part involves using Stitch to identify potential abstractions from the solutions generated in the wake phase.
For Stitch to identify and utilize abstractions effectively, programs generated by the bottom-up synthesis need to be parsed into a Lisp-like lambda calculus syntax using the de Bruijn notation~\cite{de1994survey}. 
This notation enables the nesting and composition of multiple functions with integer indices representing the binding depth of lambda expressions.
For instance, $\lambda x.(\: \lambda y.(\: x\:+\: y))$ is rewritten as $(\text{lam}\: (\text{lam}\: (+\:\$1\:\$0)))$.
Once an abstraction is identified by Stitch, it is converted back into the original format used by \acs{AB}.
Since \acs{AB} uses a bottom-up approach, it requires that abstractions are directly executable.
However, Stitch sometimes creates partial abstractions that are not immediately executable. 
To resolve this, holes in partial abstractions are replaced with variables, allowing these abstractions to be executed when given the appropriate input.
These variables are treated as additional arguments for the operation turning partial abstractions into "higher-order abstractions".
Ultimately, abstractions devoid of variables are stored as constants and are ready for execution.
To avoid unnecessary complexity, trivial abstractions, such as $\text{Add}(x2, x1)$, are not added to the \ac{DSL} but must include at least two \ac{DSL} expressions that are not variables, i.e., operations, abstractions, and/or constants.
Furthermore, abstractions must occur in two or more tasks, ensuring that abstractions are generalizable and not overly specific to individual tasks. 

In the second part, the synthesis model is adapted and trained to learn to use these newly discovered abstractions. 
The Argument Selector Module is a collection of operation-specific LSTM networks trained to propose the most suitable arguments for each operation based on the search context.
In fact, only this module needs to be updated, while all other modules are independent of the \ac{DSL} and thus continue the training in the next iteration.
When new abstractions are added to the \ac{DSL}, the Argument Selector Module must be updated to incorporate new LSTMs tailored to these operations. 
These new networks are initialized with the weights of the outermost operation used in the abstraction, providing a solid foundation for the subsequent training process.

In contrast to DreamCoder, which trains on replayed experiences from the wake phase and randomly generated tasks, \acs{AB} solely trains on \textit{task traces} constructed from the current \ac{DSL}.
To generate the task traces, an exhaustive bottom-up search using the current \ac{DSL} is performed on random tasks. 
Unlike \acs{LB}, where traces can be pre-generated through exhaustive bottom-up searches without neural guidance, \acs{AB} requires generating new training data at the beginning of each iteration to incorporate new abstractions into the training data. 
By training on randomly generated traces, \acs{AB} dynamically adapts to the evolving \ac{DSL}, enabling it to solve progressively more sophisticated problems as the language evolves.
Despite the computational overhead associated with repeated training data generation, these updates enable \acs{AB} to effectively manage the combinatorial explosion inherent in the search space.

\section{Evaluation}
\label{sec:Results}

\subsection{Experimental setup}
\paragraph{Dataset \& \ac{DSL}}
The same datasets and \ac{DSL} as in \acs{LB}~\cite{shi2024lambdabeam} were used for evaluation. 
The dataset consists of 200 tasks, split evenly between 100 handwritten and 100 synthetic tasks.
The handwritten tasks were manually crafted to cover a range of scenarios, while the synthetic tasks were generated through exhaustive bottom-up searches.
This synthetic dataset allows us to test our hypothesis that abstractions are beneficial only if they accurately reflect the domain.
In both datasets, each task involves between one and three input variables and includes three to five \ac{io} examples to guide the synthesis process.
Using the same dataset and \acs{DSL} as in \cite{shi2024lambdabeam} ensures that comparisons between the two are consistent and meaningful.
Furthermore, it allows us to directly measure the improvements brought by AbstractBeam without introducing additional variance like different domains that \acs{LB} was not yet evaluated on.
Additionally, we can evaluate the potential of Library Learning in domains not specifically designed for abstraction discovery.
Exemplary tasks and further details about the \ac{DSL} are provided in the supplementary materials.

\paragraph{Configurations}
To manage the volume and complexity of training data, we set a 1000-second time limit on each generation episode. 
Within this period, programs with a maximum length of 15 were created.
To expedite the process, up to 300 searches were conducted in parallel to generate training data. 
The data generation timeout was increased by 100 seconds for each added abstraction.
This incremental approach was necessary because generating training data involves a complete bottom-up search, and larger \acp{DSL} lead to an exponential increase in search duration.
During the wake phase, a time limit of 100 seconds per task was imposed, with random restarts occurring every 10 seconds to explore diverse solutions. 
Both algorithms were trained for a maximum of 10,000 steps per iteration. 
These parameters were chosen to be lower than those used in \acs{LB}~\cite{shi2024lambdabeam} to account for our limited computational resources. Other than that, the same hyperparameters were used.
Both approaches, \acs{AB} and \acs{LB}, were trained for 10 iterations using four RTX 2080 Ti GPUs.
The models that performed best on the handwritten training tasks were selected for evaluation. 
\acs{LB} was also trained for 10 iterations.
Yet, as the \ac{DSL} does not change in \acs{LB}, all neural modules in \acs{LB} remain unchanged and the training continues after generating more data.
This training setup was used to ensure a fair comparison and to prevent any adverse effects that might disadvantage \acs{LB}.
We used 50\% of the handwritten tasks to define abstractions.
It is important to note that these tasks were provided to the model only during the wake phase to potentially find solutions and define abstractions; they were not used to train the model itself.
The remaining 50\% were used to evaluate the performance.
To mitigate any distribution effects, a two-fold cross-validation was performed.
Each test set was evaluated five times, and Student's t-tests with a significance level of 5\% were conducted to identify significant differences between the two approaches, following checks for equal variances and normal distributions. 
The 95\% confidence intervals were calculated across trials and are displayed as error bars to convey the precision of the estimates.

\subsection{Results}
\paragraph{Success rate}
\acs{LB} solved $50.2\% \pm 2.0\%$ of all handwritten test tasks when trained on the first fold and $36.0\% \pm 2.5\%$ on the second fold.
\acs{AB} statistically significant outperformed \acs{LB} achieving $53.6\% \pm 2.3\%$ ($p < 0.05$) on the first and $39.6\% \pm 1.5\%$ ($p < 0.05$) on the second fold.
This confirms our first hypothesis that Library Learning leads to an improved performance of \acs{LB}'s synthesis.
Exemplary task solutions and a list of all found abstractions can be found in the supplementary materials.
\begin{figure}[t]
    \centering
    \begin{subfigure}{0.9\linewidth}
       \centering
       \includegraphics[width=\linewidth]{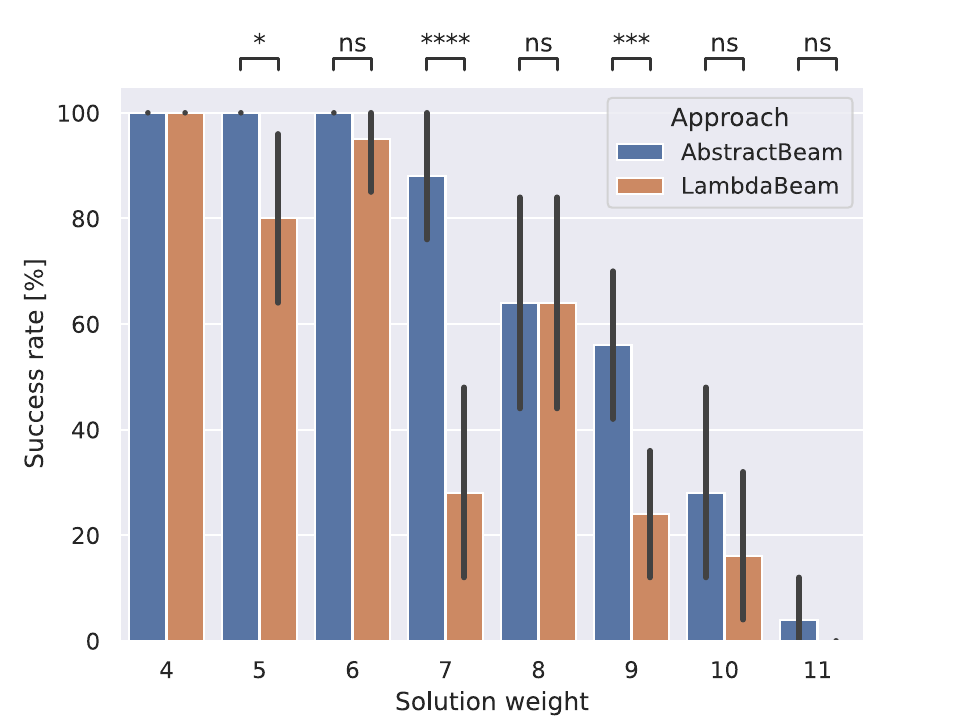}
       \caption{Success rate per program length}
       \label{fig:performance_programlength_handwritten}
    \end{subfigure}
    \begin{subfigure}{0.9\linewidth}
        \centering
        \includegraphics[width=\linewidth]{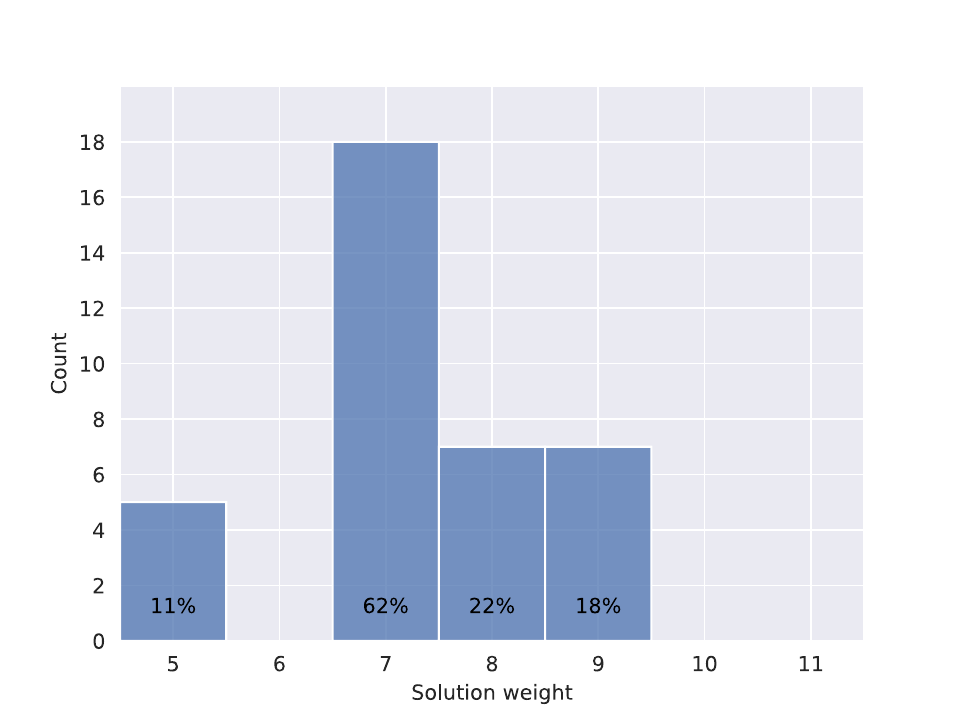}
        \caption{Abstraction usage per program length}
        \label{fig:performance_abstractionusage_handwritten}
    \end{subfigure}
    \caption[]{\acs{AB} outperforms \acs{LB}. The difference is especially dominant for program length 7. Here, 62\% and of programs contain abstractions. The differences are all significant for program lengths that use abstractions except length 8. ${ns: p >= 0.05}, {*: p < 0.05}, {{***}: p < 0.001},\\ {{****}: p < 0.0001}$}
    \label{fig:performance_abstractionimpact_fold1}
\end{figure}
As depicted in Figure~\ref{fig:performance_abstractionimpact_fold1}, significant performance differences can only be found for program lengths that use abstractions.
For program lengths where no or only a few program solutions contain abstractions, the performance does not differ statistically significantly between both approaches.
This also holds true for the second fold (figures can be found in the supplementary materials).
This further confirms that the overall performance boost is caused by the use of abstractions.

\paragraph{Efficiency}
We hypothesized that the abstraction-induced search depth reduction causes the performance boost and addresses \acs{LB}'s inefficiency in terms of time.
\begin{figure}[t!]
    \centering
    \begin{subfigure}{.89\linewidth}
        \centering
        \includegraphics[width=\linewidth]{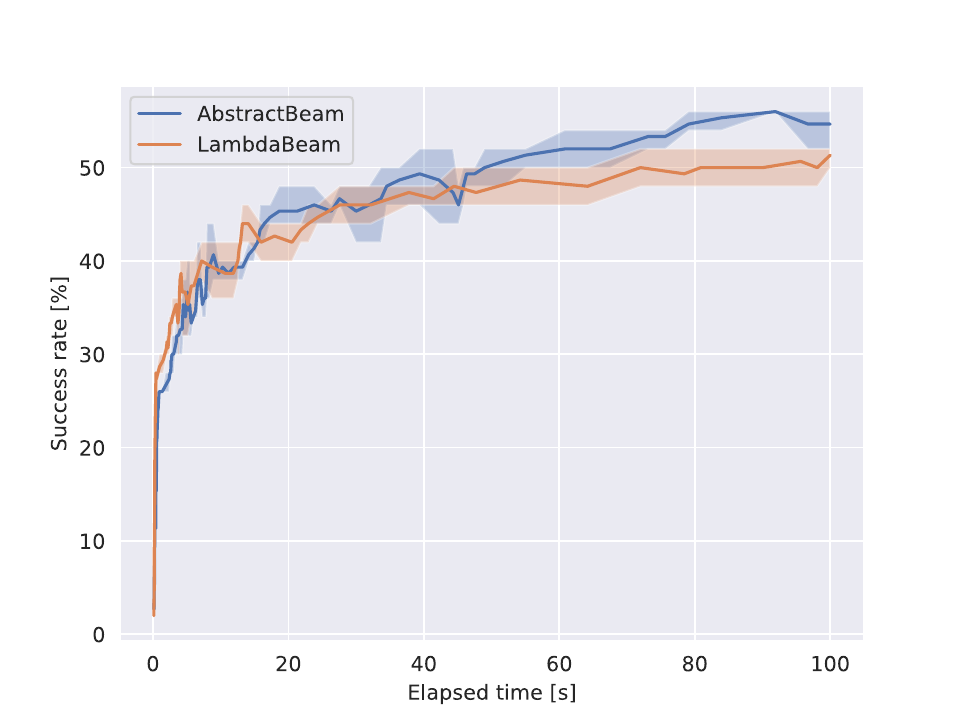}
        \caption{Enumeration time}
        \label{fig:performance_time_handwritten}
    \end{subfigure}
    \begin{subfigure}{.89\linewidth}
        \centering
        \includegraphics[width=\linewidth]{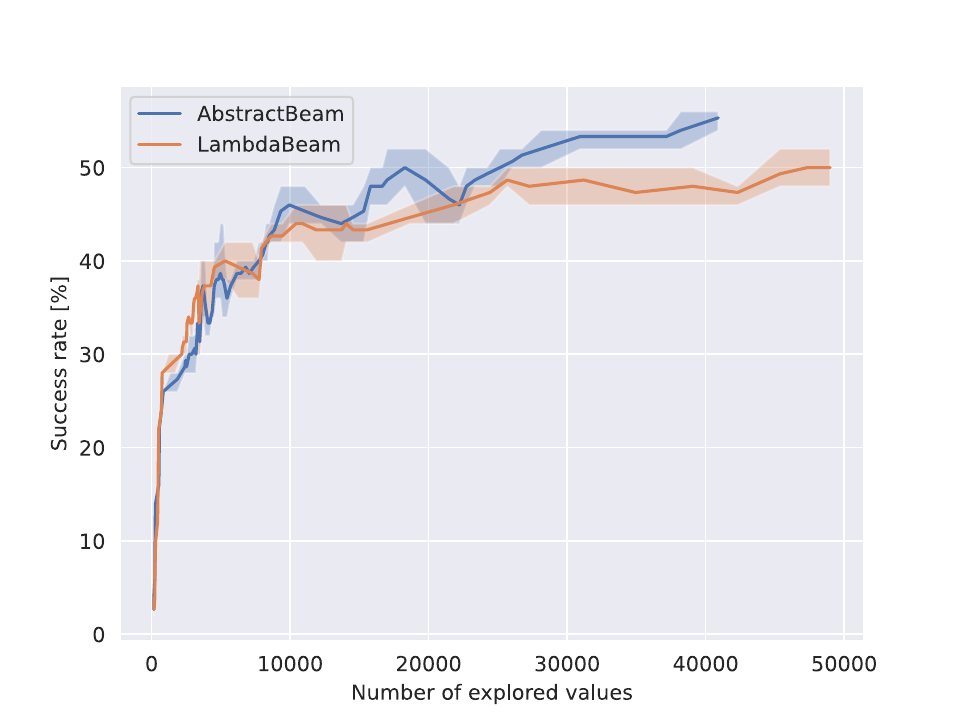}
        \caption{Candidate programs}
        \label{fig:performance_candidates_handwritten}
    \end{subfigure}
    \begin{subfigure}{.89\linewidth}
        \centering
        \includegraphics[width=\linewidth]{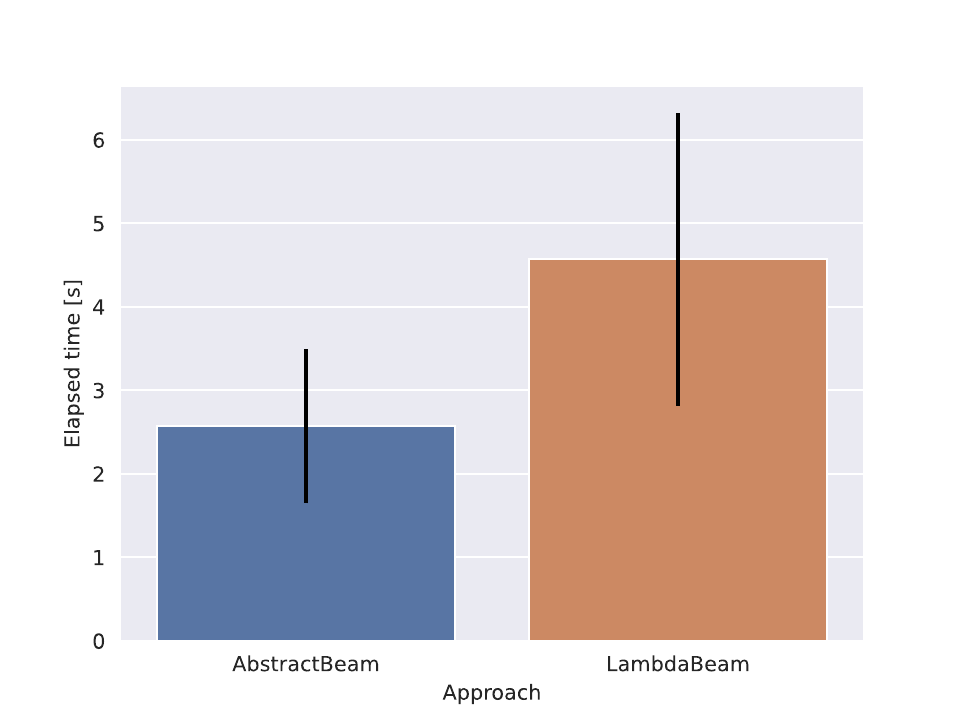}
        \caption{Enumeration times for tasks solved by both approaches}
        \label{fig:time_handwritten_2}
    \end{subfigure}
    \caption[]{With less enumeration time and fewer candidate programs \acs{AB} achieves better performance than \acs{LB}.}
    \label{fig:searchspace_reduction}
\end{figure}
To further analyze this, Figure~\ref{fig:performance_time_handwritten} shows the performance of both approaches over the enumeration time, i.e., the elapsed time until a solution was found.
Clearly, \acs{AB} shows an increased performance with less enumeration time. 
This proves the reduced search space in \acs{AB}.
Furthermore, we compared the elapsed time on handwritten tasks that were successfully solved by both \acs{AB} ($2.57s \pm 5.58s$) and \acs{LB} ($4.57s \pm 7.07s$) across all five trials.
As depicted in Figure~\ref{fig:time_handwritten_2}, \acs{AB} consistently solved these tasks in less time than \acs{LB}.
This decrease in search time demonstrates that the search space reduction facilitated by the abstractions in \acs{AB} not only increases the overall performance by solving more tasks but also enhances efficiency by clearly reducing the enumeration time required to find solutions. 
This improvement underscores the effectiveness of incorporating Library Learning and abstraction mechanisms in optimizing the search process.
Moreover, \acs{AB} also shows an improved performance with fewer candidate programs (Figure~\ref{fig:performance_candidates_handwritten}).
This also relates to the reduction of the search space depth and boosts \acs{AB}'s proficiency even further.
Interestingly, \acs{AB} enumerates fewer programs than \acs{LB} even though both algorithms used the same timeout to generate programs.
This relates to the fact that adding abstractions to the \ac{DSL} increases the search width but lowers the search depth.
With a reduced search depth, fewer candidate programs are suggested. 
As \acs{AB} still outperforms \acs{LB}, the balance between a larger \ac{DSL} and a reduced search depth is advantageous.
However, adding abstractions that are not used and bloat the \ac{DSL} will shift this balance and neglect all benefits of Library Learning.
This highlights the care that must be taken when designing the initial and altering the current \ac{DSL}.

\paragraph{Synthetic dataset}
\begin{figure}[t!]
    \centering
    \begin{subfigure}{.9\linewidth}
       \centering
       \includegraphics[width=\linewidth]{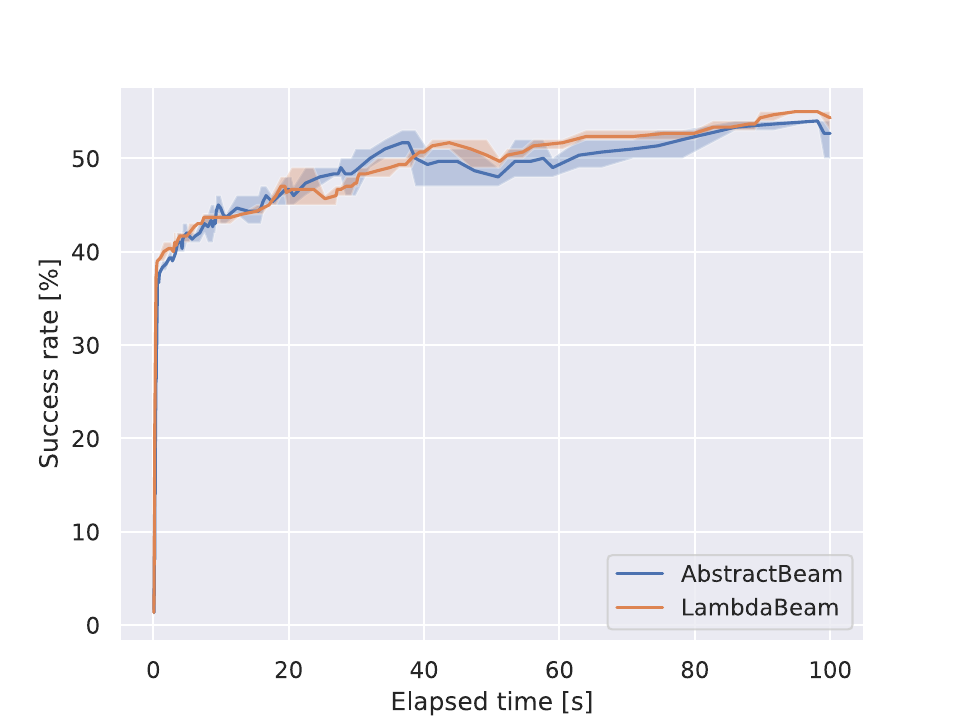}
       \caption{Enumeration time}
    \end{subfigure}
    \begin{subfigure}{.9\linewidth}
        \centering
        \includegraphics[width=\linewidth]{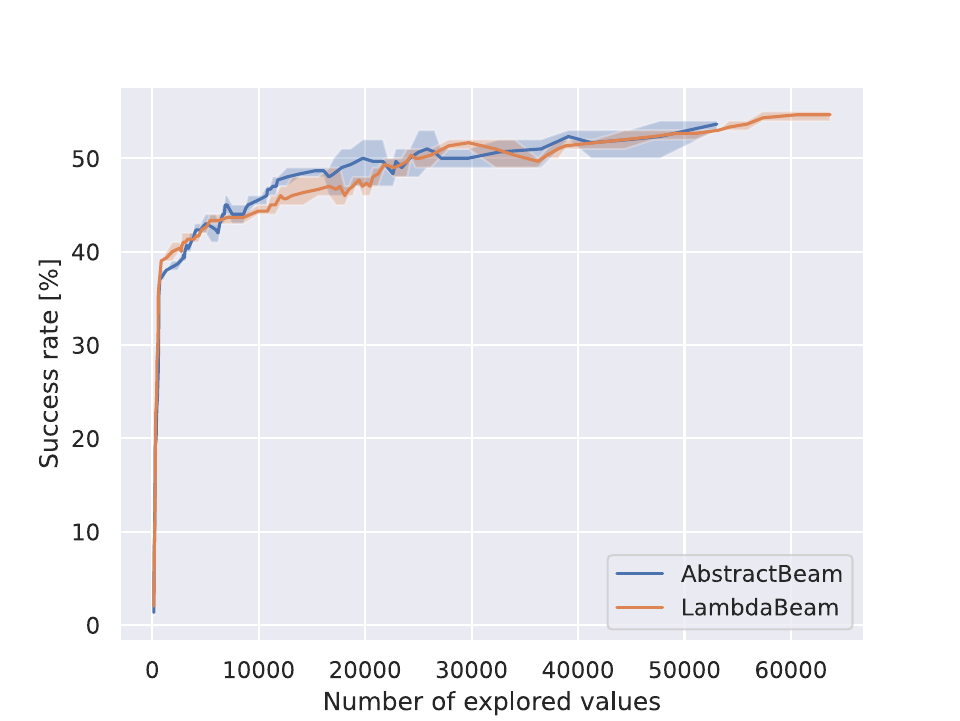}
        \caption{Candidate programs}
    \end{subfigure}
    \caption[]{Due to the distribution shift, no performance differences between \acs{LB} and \acs{AB} exist on the synthetic dataset.}
    \label{fig:performance_abstractionimpact_synthetic}
\end{figure}
To assess the degree to which abstractions scale beyond the domain they were found in, we evaluated both approaches on the synthetic dataset.
No significant performance differences were found between \acs{AB} (fold 1: $52.6\% \pm 1.4\%$, fold 2: $52.4\% \pm 1.9\%$) and \acs{LB} (fold 1: ${53.8\% \pm 1.2\%}$, fold 2: $51.2\% \pm 1.7\%$).
As shown in Figure~\ref{fig:performance_abstractionimpact_synthetic}, there are also no differences in the enumeration time and the number of evaluated candidate programs.
Consequently, even though abstractions are used in the solution of synthetic tasks, they do not reduce the search depth.
This is further supported by the enumeration times for the synthetic tasks solved by both approaches.
\acs{AB} took clearly longer to find a solution, with an average time of $3.62s \pm 12.14s$, compared to \acs{LB}, which averaged $1.37s \pm, 5.34s$.
The reason for this is that abstractions are derived from the handwritten dataset.
The handwritten dataset however does not reflect the distribution of the synthetic dataset (see the figures in the supplemental materials).
Therefore, this aligns with our hypothesis that abstractions do not generalize to out-of-distribution domains.

\paragraph{Limitations}
The key limitation of \acs{AB} is the structure of the domain which imposes two constraints:
First, Library Learning only works if abstractions are possible. 
This requires that sufficient program repetitions must exist in the solution space and sufficient tasks (or data samples) must contain these program repetitions.
Second, on a more practical note, the model must be able to solve sufficient tasks, especially in the first iteration.
Otherwise, the Library Learning algorithm cannot identify program repetitions even though sufficient tasks exist that contain a potential abstraction.
While both points are restrictions of the task domain, the latter also relates to the design of the initial \ac{DSL} that must be specific enough to solve tasks but general enough to allow repetitions.
Besides limitations imposed by Library Learning, one has to consider the overhead of regenerating training data.
Moreover, there are only slight absolute performance differences between \acs{LB} and \acs{AB} that are likely caused by the structure of the \acs{LB} domain.
Yet, our experiments prove that the statistically significant increase in solved tasks and the more efficient solution generation are attributed to the automatic introduction of domain-specific abstractions. 
In addition, our results underline the general applicability of Library Learning outside of \acp{DSL} and domains designed to show the effect of Library Learning.

\section{Related work}
Human behavior has been a fundamental source of inspiration for developing new approaches in machine learning, ranging from the creation of the first \acp{ANN}~\cite{krogh2008artificial} to the development of generative agents that imitate human beings in social interactions~\cite{pearce2023imitating}.
In the field of program synthesis, this idea is also prevalent.
Some approaches train \acp{ANN} on solution traces gathered from human problem solvers~\cite{park2023unraveling}, while others mimic human strategies by employing hierarchical approaches that break down complex problems into smaller, more manageable tasks~\cite{murali2017neural, nye2019learning, hong2021latent, shrivastava2021learning, zhong2023hierarchical, tan2023large, zelikman2023parsel}. 

These techniques aim to develop new solution strategies but often do not directly address the primary bottleneck of \acs{PS}, which is the exploding search space. 
A common approach to mitigating this issue is guiding a search algorithm toward more promising regions.
Methods for guidance vary, from leveraging prior knowledge~\cite{yin2017syntactic} to executing intermediate programs that condition subsequent searches~\cite{chen2018execution, zohar2018automatic, ellis2019write, chen2021latent, odena2020bustle}.
Generally, \acp{ANN} are employed to identify auspicious areas for exploration~\cite{yin2017syntactic, ellis2019write, lee2018accelerating}.

Our approach uses a framework inspired by DreamCoder~\cite{ellis2021dreamcoder} yet builds upon \acs{LB}'s~\cite{shi2024lambdabeam} search strategy.  
\acs{LB} itself adapts \acs{CB}~\cite{shi2022crossbeam} to allow the use of higher-order functions.
Both, employ multi-level, neural, and execution-guided synthesis.
By integrating \acs{LB} into an architecture inspired by DreamCoder we add Library Learning as a fourth strategy to reduce the search space.
Opposed to \acs{LB}, DreamCoder employs a top-down search strategy and does not condition its program generation on intermediate execution information.
Moreover, the DreamCoder \ac{DSL} consists of primitive operations yielding many commonly used subprograms. 
Furthermore, DreamCoder's list domain is manually designed to show the effect of \ac{DSL} enhancement by consisting of task clusters differing in one parameter only, e.g., \textit{add-k} with \textit{k} ranging from zero to five. 
Similar to \acs{AB}, LILO~\cite{grand2023lilo} builds upon the DreamCoder wake-sleep cycle and also uses Stitch.
This approach demonstrates that leveraging human-readable language instructions for added abstractions can improve \ac{LLM}-based \acs{PS} and alignment with human reasoning. 
This diverges from the goals of our work that focuses on creating a more resource-efficient program synthesis approach.

\section{Conclusion}
We introduce \acs{AB}, an execution-guided \acs{PS} framework that employs Library Learning to enhance its \ac{DSL} in a data-driven manner. 
By tailoring the \ac{DSL} to the specific domain it is applied to, \acs{AB} achieves improved performance and more efficient program generation.
Our results demonstrate that the use of Library Learning enables \acs{AB} to automatically identify and incorporate useful abstractions, leading to significant gains in efficiency.

However, as Library Learning is inherently data-driven, the abstractions it discovers do not generalize beyond the domain they are derived from. 
This limitation is evident in our experimental results and highlights the ongoing challenge of designing an effective \ac{DSL} that can accommodate a wide range of tasks.
Incorporating domain-specific abstractions into the \ac{DSL} likely embeds valuable domain knowledge, potentially aiding the model in better understanding the domain. 
Nonetheless, the extent to which this supports program synthesis remains an open question and warrants further investigation.

\bibliography{aaai25}

\begin{thebibliography}{39}
\providecommand{\natexlab}[1]{#1}

\bibitem[{Allamanis, Brockschmidt, and Khademi(2017)}]{allamanis2017learning}
Allamanis, M.; Brockschmidt, M.; and Khademi, M. 2017.
\newblock Learning to represent programs with graphs.
\newblock \emph{arXiv preprint arXiv:1711.00740}.

\bibitem[{Alon et~al.(2020)Alon, Sadaka, Levy, and Yahav}]{alon2020structural}
Alon, U.; Sadaka, R.; Levy, O.; and Yahav, E. 2020.
\newblock Structural language models of code.
\newblock In \emph{International conference on machine learning}, 245--256. PMLR.

\bibitem[{Balog et~al.(2016)Balog, Gaunt, Brockschmidt, Nowozin, and Tarlow}]{balog2016deepcoder}
Balog, M.; Gaunt, A.~L.; Brockschmidt, M.; Nowozin, S.; and Tarlow, D. 2016.
\newblock Deepcoder: Learning to write programs.
\newblock \emph{arXiv preprint arXiv:1611.01989}.

\bibitem[{Barke, Peleg, and Polikarpova(2020)}]{barke2020just}
Barke, S.; Peleg, H.; and Polikarpova, N. 2020.
\newblock Just-in-time learning for bottom-up enumerative synthesis.
\newblock \emph{Proceedings of the ACM on Programming Languages}, 4(OOPSLA): 1--29.

\bibitem[{Bowers et~al.(2023)Bowers, Olausson, Wong, Grand, Tenenbaum, Ellis, and Solar-Lezama}]{bowers2023top}
Bowers, M.; Olausson, T.~X.; Wong, L.; Grand, G.; Tenenbaum, J.~B.; Ellis, K.; and Solar-Lezama, A. 2023.
\newblock Top-down synthesis for library learning.
\newblock \emph{Proceedings of the ACM on Programming Languages}, 7(POPL): 1182--1213.

\bibitem[{Chen et~al.(2021)Chen, Tworek, Jun, Yuan, Pinto, Kaplan, Edwards, Burda, Joseph, Brockman et~al.}]{chen2021evaluating}
Chen, M.; Tworek, J.; Jun, H.; Yuan, Q.; Pinto, H. P. d.~O.; Kaplan, J.; Edwards, H.; Burda, Y.; Joseph, N.; Brockman, G.; et~al. 2021.
\newblock Evaluating large language models trained on code.
\newblock \emph{arXiv preprint arXiv:2107.03374}.

\bibitem[{Chen, Liu, and Song(2018)}]{chen2018execution}
Chen, X.; Liu, C.; and Song, D. 2018.
\newblock Execution-guided neural program synthesis.
\newblock In \emph{International Conference on Learning Representations}.

\bibitem[{Chen, Song, and Tian(2021)}]{chen2021latent}
Chen, X.; Song, D.; and Tian, Y. 2021.
\newblock Latent execution for neural program synthesis.
\newblock \emph{arXiv preprint arXiv:2107.00101}.

\bibitem[{Church(1985)}]{church1985calculi}
Church, A. 1985.
\newblock \emph{The calculi of lambda-conversion}.
\newblock 6. Princeton University Press.

\bibitem[{Cormen et~al.(2001)Cormen, Leiserson, Rivest, and Stein}]{cormen2001advanced}
Cormen, T.; Leiserson, C.; Rivest, R.; and Stein, C. 2001.
\newblock \emph{Advanced Algorithms}.
\newblock Citeseer.

\bibitem[{De~Bruijn(1994)}]{de1994survey}
De~Bruijn, N.~G. 1994.
\newblock A survey of the project AUTOMATH.
\newblock In \emph{Studies in Logic and the Foundations of Mathematics}, volume 133, 141--161. Elsevier.

\bibitem[{Ellis et~al.(2019)Ellis, Nye, Pu, Sosa, Tenenbaum, and Solar-Lezama}]{ellis2019write}
Ellis, K.; Nye, M.; Pu, Y.; Sosa, F.; Tenenbaum, J.; and Solar-Lezama, A. 2019.
\newblock Write, execute, assess: Program synthesis with a repl.
\newblock \emph{Advances in Neural Information Processing Systems}, 32.

\bibitem[{Ellis et~al.(2021)Ellis, Wong, Nye, Sabl{\'e}-Meyer, Morales, Hewitt, Cary, Solar-Lezama, and Tenenbaum}]{ellis2021dreamcoder}
Ellis, K.; Wong, C.; Nye, M.; Sabl{\'e}-Meyer, M.; Morales, L.; Hewitt, L.; Cary, L.; Solar-Lezama, A.; and Tenenbaum, J.~B. 2021.
\newblock Dreamcoder: Bootstrapping inductive program synthesis with wake-sleep library learning.
\newblock In \emph{Proceedings of the 42nd acm sigplan international conference on programming language design and implementation}, 835--850.

\bibitem[{Grand et~al.(2023)Grand, Wong, Bowers, Olausson, Liu, Tenenbaum, and Andreas}]{grand2023lilo}
Grand, G.; Wong, L.; Bowers, M.; Olausson, T.~X.; Liu, M.; Tenenbaum, J.~B.; and Andreas, J. 2023.
\newblock Lilo: Learning interpretable libraries by compressing and documenting code.
\newblock \emph{arXiv preprint arXiv:2310.19791}.

\bibitem[{Gulwani et~al.(2017)Gulwani, Polozov, Singh et~al.}]{gulwani2017program}
Gulwani, S.; Polozov, O.; Singh, R.; et~al. 2017.
\newblock Program synthesis.
\newblock \emph{Foundations and Trends{\textregistered} in Programming Languages}, 4(1-2): 1--119.

\bibitem[{Hong et~al.(2021)Hong, Dohan, Singh, Sutton, and Zaheer}]{hong2021latent}
Hong, J.; Dohan, D.; Singh, R.; Sutton, C.; and Zaheer, M. 2021.
\newblock Latent programmer: Discrete latent codes for program synthesis.
\newblock In \emph{International Conference on Machine Learning}, 4308--4318. PMLR.

\bibitem[{Krogh(2008)}]{krogh2008artificial}
Krogh, A. 2008.
\newblock What are artificial neural networks?
\newblock \emph{Nature biotechnology}, 26(2): 195--197.

\bibitem[{Lee et~al.(2018)Lee, Heo, Alur, and Naik}]{lee2018accelerating}
Lee, W.; Heo, K.; Alur, R.; and Naik, M. 2018.
\newblock Accelerating search-based program synthesis using learned probabilistic models.
\newblock \emph{ACM SIGPLAN Notices}, 53(4): 436--449.

\bibitem[{Murali et~al.(2017)Murali, Qi, Chaudhuri, and Jermaine}]{murali2017neural}
Murali, V.; Qi, L.; Chaudhuri, S.; and Jermaine, C. 2017.
\newblock Neural sketch learning for conditional program generation.
\newblock \emph{arXiv preprint arXiv:1703.05698}.

\bibitem[{Nye et~al.(2019)Nye, Hewitt, Tenenbaum, and Solar-Lezama}]{nye2019learning}
Nye, M.; Hewitt, L.; Tenenbaum, J.; and Solar-Lezama, A. 2019.
\newblock Learning to infer program sketches.
\newblock In \emph{International Conference on Machine Learning}, 4861--4870. PMLR.

\bibitem[{Odena et~al.(2020)Odena, Shi, Bieber, Singh, Sutton, and Dai}]{odena2020bustle}
Odena, A.; Shi, K.; Bieber, D.; Singh, R.; Sutton, C.; and Dai, H. 2020.
\newblock BUSTLE: bottom-up program synthesis through learning-guided exploration.
\newblock \emph{arXiv preprint arXiv:2007.14381}.

\bibitem[{Park et~al.(2023)Park, Im, Hwang, Lim, Ualibekova, Kim, and Kim}]{park2023unraveling}
Park, J.; Im, J.; Hwang, S.; Lim, M.; Ualibekova, S.; Kim, S.; and Kim, S. 2023.
\newblock Unraveling the arc puzzle: Mimicking human solutions with object-centric decision transformer.
\newblock \emph{arXiv preprint arXiv:2306.08204}.

\bibitem[{Pea and Kurland(1984)}]{pea1984cognitive}
Pea, R.~D.; and Kurland, D.~M. 1984.
\newblock On the cognitive effects of learning computer programming.
\newblock \emph{New ideas in psychology}, 2(2): 137--168.

\bibitem[{Pearce et~al.(2023)Pearce, Rashid, Kanervisto, Bignell, Sun, Georgescu, Macua, Tan, Momennejad, Hofmann et~al.}]{pearce2023imitating}
Pearce, T.; Rashid, T.; Kanervisto, A.; Bignell, D.; Sun, M.; Georgescu, R.; Macua, S.~V.; Tan, S.~Z.; Momennejad, I.; Hofmann, K.; et~al. 2023.
\newblock Imitating human behaviour with diffusion models.
\newblock \emph{arXiv preprint arXiv:2301.10677}.

\bibitem[{Pierce(2002)}]{pierce2002types}
Pierce, B.~C. 2002.
\newblock \emph{Types and programming languages}.
\newblock MIT press.

\bibitem[{Polozov and Gulwani(2015)}]{polozov2015flashmeta}
Polozov, O.; and Gulwani, S. 2015.
\newblock Flashmeta: A framework for inductive program synthesis.
\newblock In \emph{Proceedings of the 2015 ACM SIGPLAN International Conference on Object-Oriented Programming, Systems, Languages, and Applications}, 107--126.

\bibitem[{Shi, Bieber, and Singh(2022)}]{shi2022tf}
Shi, K.; Bieber, D.; and Singh, R. 2022.
\newblock Tf-coder: Program synthesis for tensor manipulations.
\newblock \emph{ACM Transactions on Programming Languages and Systems (TOPLAS)}, 44(2): 1--36.

\bibitem[{Shi, Bieber, and Sutton(2020)}]{shi2020incremental}
Shi, K.; Bieber, D.; and Sutton, C. 2020.
\newblock Incremental sampling without replacement for sequence models.
\newblock In \emph{International Conference on Machine Learning}, 8785--8795. PMLR.

\bibitem[{Shi et~al.(2022)Shi, Dai, Ellis, and Sutton}]{shi2022crossbeam}
Shi, K.; Dai, H.; Ellis, K.; and Sutton, C. 2022.
\newblock CrossBeam: Learning to search in bottom-up program synthesis.
\newblock \emph{arXiv preprint arXiv:2203.10452}.

\bibitem[{Shi et~al.(2024)Shi, Dai, Li, Ellis, and Sutton}]{shi2024lambdabeam}
Shi, K.; Dai, H.; Li, W.-D.; Ellis, K.; and Sutton, C. 2024.
\newblock LambdaBeam: Neural Program Search with Higher-Order Functions and Lambdas.
\newblock \emph{Advances in Neural Information Processing Systems}, 36.

\bibitem[{Shi, Steinhardt, and Liang(2019)}]{shi2019frangel}
Shi, K.; Steinhardt, J.; and Liang, P. 2019.
\newblock Frangel: component-based synthesis with control structures.
\newblock \emph{Proceedings of the ACM on Programming Languages}, 3(POPL): 1--29.

\bibitem[{Shrivastava, Larochelle, and Tarlow(2021)}]{shrivastava2021learning}
Shrivastava, D.; Larochelle, H.; and Tarlow, D. 2021.
\newblock Learning to combine per-example solutions for neural program synthesis.
\newblock \emph{Advances in Neural Information Processing Systems}, 34: 6102--6114.

\bibitem[{Stevens, Antiga, and Viehmann(2020)}]{stevens2020deep}
Stevens, E.; Antiga, L.; and Viehmann, T. 2020.
\newblock \emph{Deep learning with PyTorch}.
\newblock Manning Publications.

\bibitem[{Tan and Motani(2023)}]{tan2023large}
Tan, J. C.~M.; and Motani, M. 2023.
\newblock Large Language Model (LLM) as a System of Multiple Expert Agents: An Approach to solve the Abstraction and Reasoning Corpus (ARC) Challenge.
\newblock \emph{arXiv preprint arXiv:2310.05146}.

\bibitem[{Van~Deursen, Klint, and Visser(2000)}]{van2000domain}
Van~Deursen, A.; Klint, P.; and Visser, J. 2000.
\newblock Domain-specific languages: An annotated bibliography.
\newblock \emph{ACM Sigplan Notices}, 35(6): 26--36.

\bibitem[{Yin and Neubig(2017)}]{yin2017syntactic}
Yin, P.; and Neubig, G. 2017.
\newblock A syntactic neural model for general-purpose code generation.
\newblock \emph{arXiv preprint arXiv:1704.01696}.

\bibitem[{Zelikman et~al.(2023)Zelikman, Huang, Poesia, Goodman, and Haber}]{zelikman2023parsel}
Zelikman, E.; Huang, Q.; Poesia, G.; Goodman, N.; and Haber, N. 2023.
\newblock Parsel: Algorithmic Reasoning with Language Models by Composing Decompositions.
\newblock \emph{Advances in Neural Information Processing Systems}, 36: 31466--31523.

\bibitem[{Zhong et~al.(2023)Zhong, Lindeborg, Zhang, Lim, and Sun}]{zhong2023hierarchical}
Zhong, L.; Lindeborg, R.; Zhang, J.; Lim, J.~J.; and Sun, S.-H. 2023.
\newblock Hierarchical neural program synthesis.
\newblock \emph{arXiv preprint arXiv:2303.06018}.

\bibitem[{Zohar and Wolf(2018)}]{zohar2018automatic}
Zohar, A.; and Wolf, L. 2018.
\newblock Automatic program synthesis of long programs with a learned garbage collector.
\newblock \emph{Advances in neural information processing systems}, 31.

\end{thebibliography}

\includepdf[pages=-]{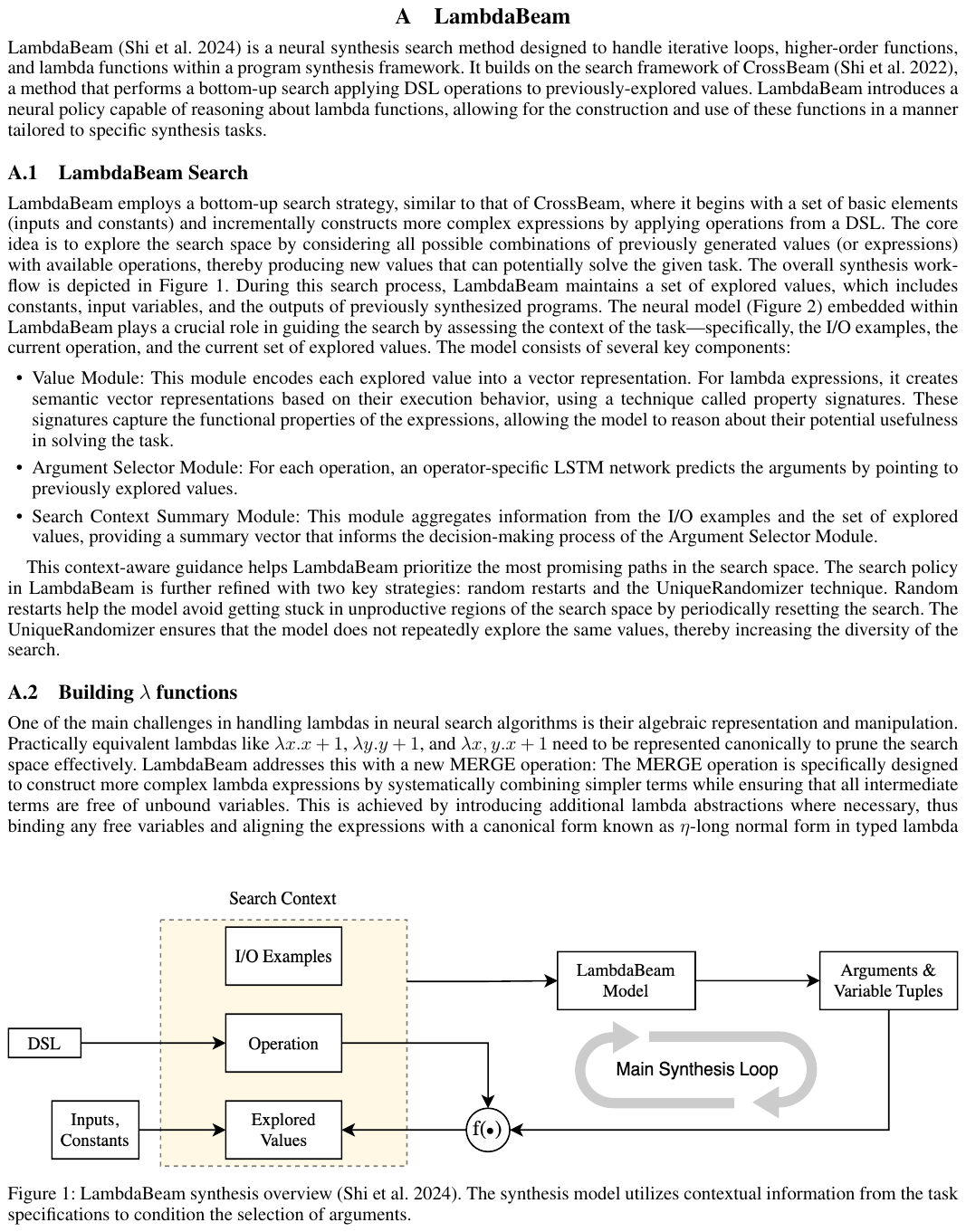}
\end{document}